\documentclass[twocolumn]{aastex61}

\newcommand{\be}{\begin{enumerate}}
\newcommand{\ee}{\end{enumerate}}

\usepackage{color}

\begin{document}

\title{Winds in Star Clusters Drive Kolmogorov Turbulence}

\correspondingauthor{Monica Gallegos-Garcia}
\email{mgallegosgarcia@u.northwestern.edu}

\author{Monica Gallegos-Garcia}
\affiliation{Department of Physics and Astronomy, Northwestern University, 2145 Sheridan Road, Evanston, IL 60208, USA}

\affiliation{Center for Interdisciplinary Exploration and Research in Astrophysics (CIERA),1800 Sherman, Evanston, IL 60201, USA}

\author{Blakesley Burkhart}

\affiliation{Center for Computational Astrophysics, Flatiron Institute, 162 Fifth Avenue, New York, NY 10010, USA}
\affiliation{Department of Physics and Astronomy, Rutgers, The State University of New Jersey, 136 Frelinghuysen Rd, Piscataway, NJ 08854, USA}

\author{Anna Rosen}
\affiliation{Harvard-Smithsonian Center for Astrophysics, 60 Garden Street, Cambridge, MA 02138, USA}
\affiliation{Einstein Fellow}
\affiliation{ITC Fellow}

\author{Jill P. Naiman}
\affiliation{School of Information Sciences, University of Illinois, Urbana-Champaign, IL, 61820}

\author{Enrico Ramirez-Ruiz}
\affiliation{Department of Astronomy and Astrophysics, University of California, Santa Cruz, CA 95064, USA}
\affiliation{Niels Bohr Institute, University of Copenhagen, Blegdamsvej 17, 2100 Copenhagen, Denmark}

\begin{abstract}
Intermediate and massive stars drive fast and powerful isotropic winds that interact with the winds of nearby stars in star clusters and  the surrounding interstellar medium (ISM). Wind-ISM collisions generate astrospheres around these stars that contain hot $T\sim10^7$ K gas that adiabatically expands. As individual bubbles expand and collide they  become unstable, potentially driving turbulence in star clusters.
In this paper we use hydrodynamic simulations to model a densely populated young star cluster within a homogeneous cloud to study stellar wind collisions with the surrounding ISM. We model a mass-segregated cluster of 20 B-type young main sequence stars with masses ranging from 3--17 $M_{\odot}$. We evolve the winds for $\sim$11 kyrs and show that wind-ISM  collisions and over-lapping wind-blown bubbles around B-stars mixes the hot gas and ISM material generating Kolmogorov-like turbulence on small scales early in its evolution. We discuss how turbulence driven by stellar winds may impact the subsequent generation of star formation in the cluster.

\end{abstract}
\keywords{ISM: Turbulence -- stars: winds  }

\section{Introduction}
\label{sec:intro}
Feedback from stellar winds play an important role in shaping the structure of the interstellar medium \citep[ISM,][]{krumreview2014}. Massive stars produce powerful winds since the mass loss rates and wind velocities are determined by the star's radiation output \citep{Castor1975}. Intermediate- and low-mass stars also contribute to producing ionized bubbles in the ISM, i.e. so-called astrospheres \citep{2004LRSP....1....2W,2016A&A...586A.114M}, which are a potential source of local ISM turbulence \citep{2017ApJ...849L..10B}, cosmic rays \citep{2015MNRAS.448..207D}, dust processing \citep{2017MNRAS.465.1573K}, and can be used to identify runaway stars \citep{Peri_2012}.

In regards to massive stars, early theoretical models by \cite{1975ApJ...200L.107C} and \cite{1977ApJ...218..377W} demonstrated that the interaction between fast, isotropic stellar winds and the surrounding ISM produces a large cavity or ``bubble'' surrounded by a thin shell of dense, cold material. In agreement with these models, parsec-scale circular cavities are regularly found in regions of high-mass star formation \citep{churchwell2006, churchwell2007, beaumontwilliams2010, deharveng2010}. 
Such features likely contribute to parsec-scale turbulence in these environments and drive density fluctuations that influence subsequent generations of stars \citep{Offner2015, Burkhart2018}.

Although it was previously thought that only winds from O or early B-type stars could drive bubbles in molecular clouds, numerous shells have been found in low- and intermediate-mass star forming regions \citep{Arce2011, 2015ApJS..219...20L}. These studies concluded that these bubbles are likely driven by stellar winds from intermediate-mass stars and the energetics of these bubbles may help sustain turbulence in the Perseus and Taurus star-forming regions, which may explain the observed density and velocity power spectrum in Perseus \citep{Pingel2018,padoan2006}.

To study the development and expansion of wind-blown bubbles around intermediate-mass stars and their contribution to sustaining turbulence in molecular clouds,  \cite{Offner2015} performed isothermal magnetohydrodynamic (MHD) simulations that modeled stellar wind momentum feedback from intermediate-mass main sequence stars embedded in a turbulent molecular cloud. Similar to \cite{Arce2011}, they find that for a random distribution of stars whose individual winds do not interact, a mass-loss rate of  $ \geq 10^{-7} M_{\odot} \rm yr^{-1}$ and a wind velocity of $200 \; \rm{km \ s^{-1}}$ is required to drive the shells observed in a Perseus-like molecular cloud. Their study also showed that the stellar winds that produce and drive the expansion of these shells do not produce clear features in the Fourier spectra of density and momentum but do impact the Fourier velocity spectrum. They conclude that stellar winds with high mass-loss rates can contribute to turbulence in molecular clouds. 

A natural extension in studying how wind-blown bubbles interact with the ISM and contribute to large-scale turbulence in molecular clouds is to study how these bubbles interact with one another in clustered environments. Expanding shells have been observed around small star clusters like the $\rho$-Oph cluster, which contains five B-stars located in the Ophiuchus molecular cloud \citep{ladalada2003, chen_inprep}. In this scenario, fast winds ejected from stars collide with winds from neighboring stars causing the bubbles to overlap and form a collective ``cluster wind'' \citep{2000ApJ...536..896C}. The resulting ``super-bubble,'' which is filled with hot and diffuse gas, eventually expands beyond the star cluster itself \citep{1980ApJ...238L..27B, 2003MNRAS.339..280S, 2007MNRAS.380.1198R,2008ApJ...684.1384R}.  Similar to the single wind-blown shell, where Rayleigh-Taylor and Kelvin-Helmholtz instabilities lead to turbulent mixing \citep{1984ApJ...278L.115M, 2006ApJS..164..477N}, wind-wind collisions in a multiple star system may also lead to instabilities within the cluster wind and produce small-scale turbulence within the ISM. This turbulent motion may act in the same way as the single star case, introducing energy and turbulence into its environment as the super-bubble grows. 

Motivated by this, in this Letter we perform hydrodynamic simulations to model the collective cluster wind from a dense star cluster of young B-type stars embedded in a uniform molecular cloud to determine how wind-wind collisions and overlapping bubbles can drive turbulence in star clusters. This is in contrast to  \cite{2000ApJ...536..896C} in which only a single mass of star was used in the cluster simulations. \cite{Offner2015} use an isothermal equation of state and therefore only follow the momentum injection by winds of young intermediate mass stars. Here we use an adiabatic equation of state and calculate the energy losses using a realistic cooling function, which allows us to fully capture the kinetic energy and momentum injection from the fast stellar winds and to follow the  expansion of the resulting super bubble. We investigate on what time scales turbulence can be effectively generated within a cluster by these intermediate- and high-mass stars. 

This Letter is organized as follows: in Section \ref{sec:methods} we describe the stellar wind properties, the initial conditions, and the corresponding physics of our simulation. In Section \ref{sec: simulation snapshots}, we describe the bulk properties of our simulations and show how overlapping wind bubbles can drive turbulence in young star clusters. In Section \ref{sec:power spectrum} we show the evolution of the density-weighted power spectrum and PDFs of physical properties of interest such as the temperature, density and Mach numbers. In Section \ref{sec:shell properties} we show the cooling efficiently of the collective cluster wind. Finally, in Section \ref{sec:discussion} we summarize our findings and discuss their implications.

\section{Methods}\label{sec:methods}

We assume a star cluster mass of $400 \; M_{\odot}$ with individual star masses chosen by stochastically sampling the Kroupa initial mass function \citep{2001MNRAS.322..231K}. We only model the 20 most massive stars in the cluster (masses ranging between $3.2$ -- $17 \; M_{\odot}$) because the energy and momentum injected by their winds dominate over the total momentum and energy of the entire stellar population in the cluster \citep{Rosen2014}. The 20 stars in our cluster are mass-segregated, with a cluster radius of $r = 0.14 \rm \ pc$ and a stellar density profile resembling the Orion Nebula Cluster \citep{2014ApJ...795...55D}. We simulate the wind-wind interactions in the star cluster for 11 kyr, up to the point where the cluster wind bubble expands to a radius of $\sim 0.22$~pc.  

For the mass range chosen, the stellar winds are radiatively driven \citep{Vink2001}. The values of the isotropic wind mass-loss ($\dot{M}$), and wind temperature ($T_{\rm w}$) are taken from the Modules for Experiments in Stellar Astrophysics (MESA) Isochrones and Stellar Tracks \citep[MIST,][]{2016ApJ...823..102C}.  Wind velocity ($v_{\rm w}$) is taken to be the escape velocity of the stars, an adequate approximation for the stars simulated here \citep{Naiman2018}. The mass-loss values range from $ \sim 2~\times~10^{-12}$ -- $2~\times~10^{-8} M_{\odot} \rm yr^{-1}$, the wind temperature range from $13$ -- $29\times 10^{3} \ \rm K$, and  the wind velocities are between $\sim 790$ -- $940 \ \rm km \ s^{-1}$. These values correspond to a young star cluster with an age between 4 -- 8$ \rm \ Myr$.

We perform our simulations with FLASH, a 3D adaptive mesh refinement grid-based hydrodynamics code, which allows us to include self-gravity and cooling \citep{2000ApJS..131..273F}. We use the default refinement criteria in FLASH with density as the refinement variable and additionally enforce maximum refinement for cells near the stars. We assume an ideal equation of state where the gas pressure is given by $P=(\gamma-1) e_{T}$, where $\rho$ is the gas density, $e_{T}$ is the thermal energy density per unit mass, and $\gamma$ is the adiabatic index, which we take to be $5/3$. We assume the energy equation is modified by a cooling rate of the form $Q(\vec{r},t)~=~n_i(\vec{r},t) n_e(\vec{r},t) \Lambda(T,Z)$.  This is derived from the electron and ion number densities, $n_e(\vec{r},t)$ and $n_i(\vec{r},t)$ and cooling function for gas of temperature $T$ and metallicity $Z$, where $\Lambda(T,Z)$ is taken from \cite{2007ApJS..168..213G} for $T>10^4 \, \rm K$ and from \cite{1972ARA&A..10..375D} for $10 \ {\rm K} \le T \le 10^4 \ {\rm K}$.  Metallicity is fixed at solar.

The star cluster modeled is embedded in a non-turbulent background so that we can self-consistently follow the driving of turbulence generated only by winds. While our initial condition of a uniform background density is certainly idealized, it is likely that turbulence is significantly damped on the scales of a few tenths of a parsec due to various viscous and MHD damping mechanisms \citep{2008ApJ...684..380L,2015ApJ...805..118B,2016ApJ...826..166X,Qian_2018}. As we are interested to study the direct impact of turbulence produced by the star cluster we restrict ourselves to a case in which the ambient medium is uniform. The ambient medium has a density of $n_{\rm amb} = 10^3 \ \rm cm^{-3}$ and a cloud temperature of $10 \rm \ K$.  The box size is (1.24 pc)$^3$ with a finest spatial resolution of $\sim 120 \rm AU$. For reference, the shell radii in Perseus identified by \cite{Arce2011} range within 0.14 -- 2.79 pc. They also use a cloud density $\sim 10^4 \ \rm cm^{-3}$ to calculate mass-loss rates of the stars embedded within the shells. \cite{chen_inprep} find an average radius of $\sim 1.36$ pc for the shell in Ophiuchus, which is likely being driven by 5 B-type stars.

To model wind feedback on the ISM by the stars, we inject the stellar wind over a spherically-symmetric sphere surrounding each star with a diameter of 16 cells (corresponding to $\sim$1000 AU in radius). This follows the results of \cite{ruffert1994} that suggests $>$8 cells are required for sink or source grid sources in hydrodynamical simulations. Within this sphere, the wind density decreases proportional to the inverse square distance from the center, while the wind velocity magnitude and temperature remain constant.


\section{Results} \label{sec:results}

\subsection{Evolution of the Wind-blown Bubbles in Star Clusters}\label{sec: simulation snapshots}

\begin{figure*}
\hbox{\hspace{4em}
\includegraphics[width=0.86\linewidth]{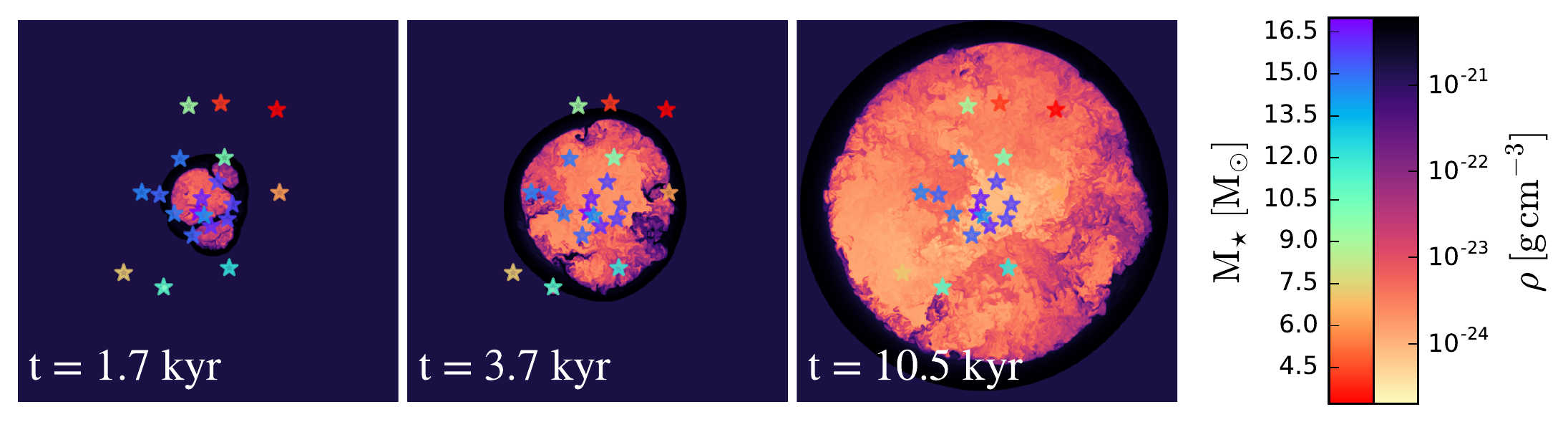}}
\hbox{\hspace{4em}
\includegraphics[width=0.846\linewidth]{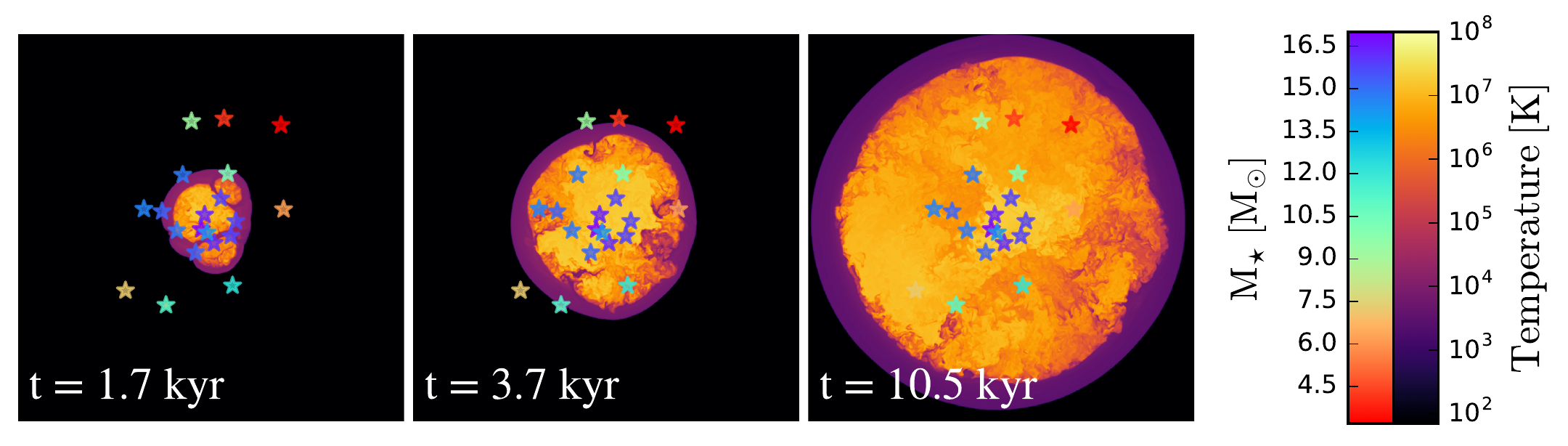}}
\hbox{\hspace{4em}
\includegraphics[width=0.85\linewidth]{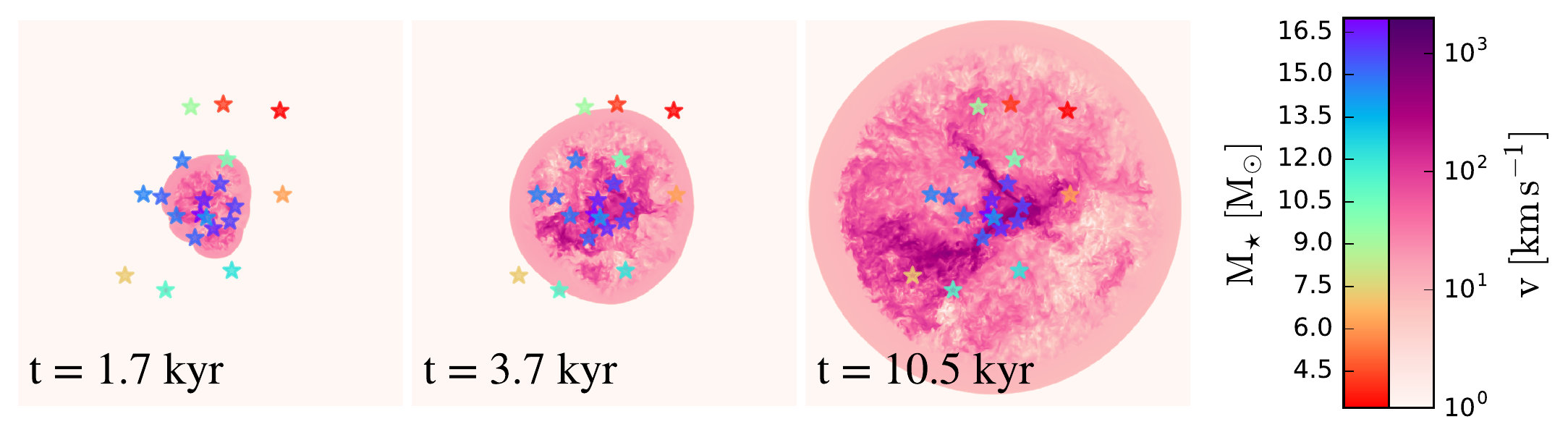}}

\caption{ Time series of slices along the y-z plane of the simulation. We show gas density (top row), gas temperature (middle row), and the velocity magnitude of the gas (bottom row). Star markers indicate the projected stellar locations onto the slice plane and denote star mass with color.  The first column shows the wind-blown bubbles before the single cluster wind is formed. The second column shows a snapshot when areas of significant mixing of material appear. The third column shows one of the final snapshot of our simulation.}
\label{fig:timeseries1}
\end{figure*}

\begin{figure*}
\centering
\includegraphics[width=1.0\linewidth]{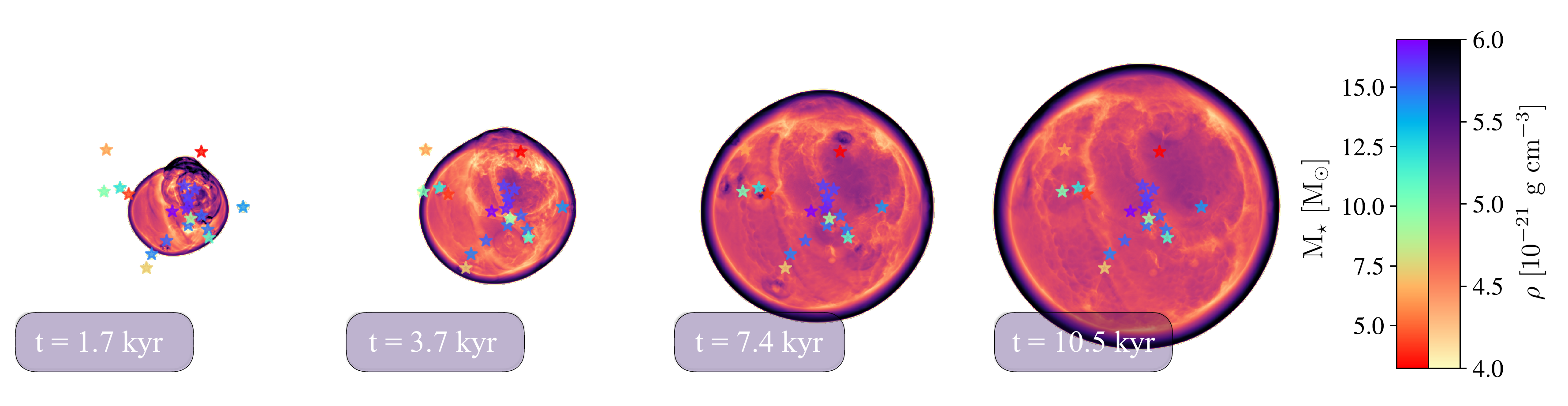}

\includegraphics[width=1.0\linewidth]{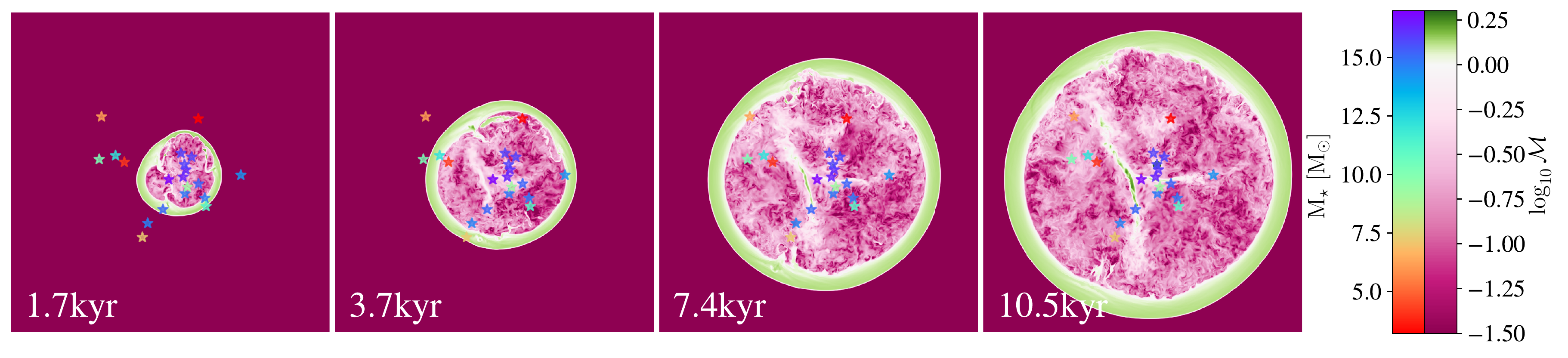}

\caption{Top row: density-weighted projections of density through the x-y plane of the simulation. Bottom row: Mach number $\mathcal{M}$ slices along the x-y plane of the simulation. Green is supersonic material and pink is subsonic. Most of the wind material is subsonic except for the outer shell. This is because the high densities in this region allow the gas to cool to lower temperatures, decreasing the speed of sound in this region. }

\label{fig:timeseries2}

\end{figure*}

Figure \ref{fig:timeseries1} shows a time series of slices through the simulation for gas density (top row), gas temperature (middle row), and the velocity magnitude of the gas (bottom row). We denote star mass and star locations projected on the slice plane with colored stars.  

The density panels in Figure \ref{fig:timeseries1} show the result of the stellar winds from the most massive stars near the center quickly colliding and merging into one large bubble, similar to those observed in molecular clouds \citep{ladalada2003, Rosen2014} and predicted by \cite{2000ApJ...536..896C}. The first panel at 1.7 kyrs shows three bubbles formed by the central most massive stars. The larger bubble deforms the spherical shape of the other two bubbles as it expands. At $t \sim$ 2 kyr the largest shell bursts along this plane when it reaches the position of another star. The three bubbles begin to coalesce with each other to form the resulting cluster wind and engulf the lower-mass stars at larger radii in the same manner. The second density panel at 3.7 kyr shows mixing of low-density stellar wind material with high-density material that was initially between the separate bubbles (see first panel). These mixing features are most prominent when the bubbles merge near the edges of the cluster wind since that is where more high-density swept-up material resides. We associate the mixing features to be turbulent instabilities that are likely a result of Kelvin-Helmholtz instabilities caused by the wind-wind collisions. In agreement with our results, \cite{Krause2013} also finds these instabilities develop at the locations of intersecting wind bubbles. Throughout the simulation mixing features are also prominent along the inner edge of the shell as high-temperature and low-density gas is pushed onto it leading to turbulent mixing at the shell-bubble boundary \citep{Rosen2014}. The last panel at 10.5 kyr shows the cluster wind expanded to a radius of $\sim$0.21 pc. During the entire evolution the gas motion is dominated by the wind feedback from the central stars. This can be seen by the fact that the central region of the cluster bubble is kept at lower densities. This is due to the high velocities of the stellar winds pushing material in this vicinity away at all times. Although self-gravity of the gas is calculated in these simulations, we find that self-gravity is dynamically unimportant within the shell since the total mass within the bubble, $M_{\rm B} \sim 0.01 \; M_{\odot}$, is low.

The second row in Figure~\ref{fig:timeseries1} shows the temperature evolution of the star cluster. As the winds collide with the surrounding ISM and other wind material, the kinetic energy is thermalized resulting in high temperatures of $T = 10^7$--$10^8$~K within the wind bubble. This hot gas adiabatically expands, which we properly account for since we use an adiabatic, rather than isothermal, equation of state. It is this adiabatic expansion that dominates the over-all expansion of the combined wind bubble \citep{2000ApJ...536..896C}.
Similar to the density panels, we can see a mixing of hot wind material with cooler shell material. Again this mixing is most prominent when the bubbles merge and at the bubble edges. The cluster gas does not cool significantly over timescales shown here. This cooling inefficiency is likely due to low cooling rates that are achieved by the low-density and high-temperature gas at solar metallicity \citep{Rosen2014}. In contrast, the high-density and low-temperature shell cools slightly over the timescale shown here, which we discuss in more detail in Section~\ref{sec:cool}.

\begin{figure*}
\centering\includegraphics[width=1.0\linewidth]{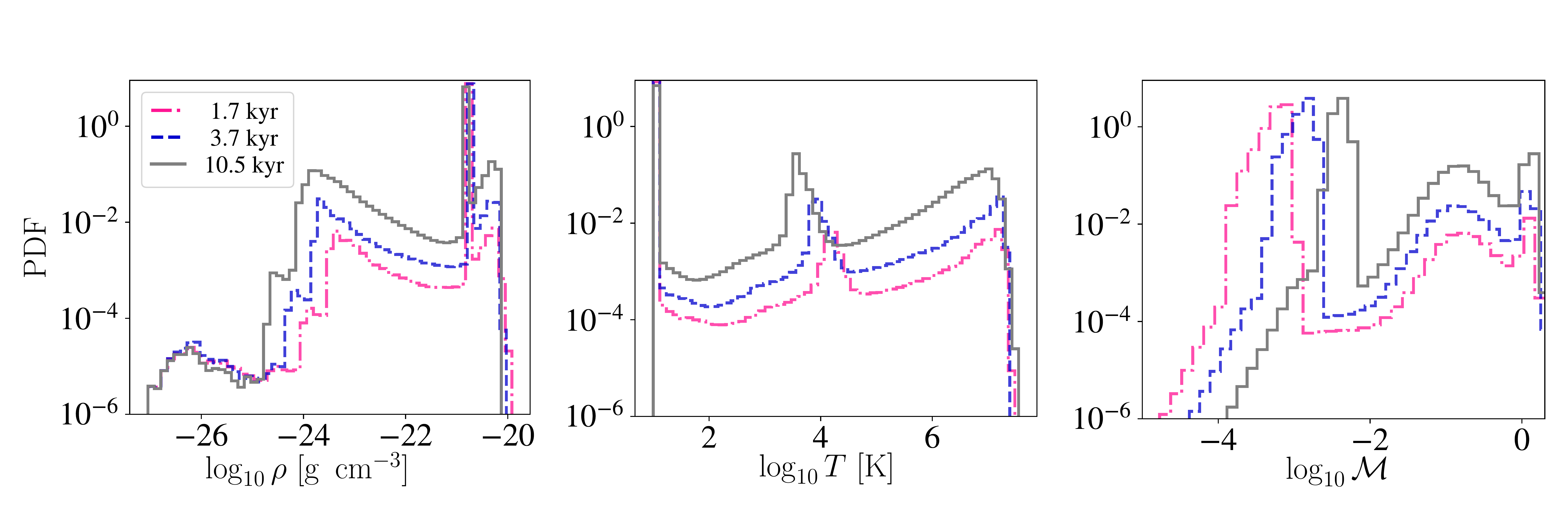}
\caption{Probability density functions for density, temperature, and Mach number at three snapshots in the simulation. The peaks at $\rho \sim 10^{21} \rm \ \ g \ cm^{-3}$, T$ \sim 10 \ \rm K$, and the peak between $ \mathcal{M} \sim 10^{-4} $--$10^{-2}$ correspond to the background material. Left: Density PDF. Middle: The pink, dotted-dashed peak at T$ \sim 10^{4} \ \rm K$ corresponds to the cluster shell at $t=1.7 \ \rm kyr$. As the simulation evolves this region cools the most.  Right: The cluster bubble is dominated by sub-sonic material. }
\label{fig:density_PDF_timeseries}
\end{figure*}

The last row of Figure \ref{fig:timeseries1} shows the velocity magnitude in the simulation. We see that the high-velocity material dominates the inner regions of the cluster wind where the density is the lowest. As the shell expands and sweeps up material from the ambient medium, it slows and cools (cooling of the shell shown in the middle panel of Figure \ref{fig:density_PDF_timeseries}).

Figure~\ref{fig:timeseries2} shows a time series of mass-weighted density projections (top row) and slices through the simulation of Mach number $\mathcal{M}$ (bottom row), the ratio of the gas velocity magnitude to the sound speed, $v/c_{\rm s}$. Pink corresponds to subsonic, $\mathcal{M}<1$,  and green corresponds to supersonic, $\mathcal{M}>1$. In the $\mathcal{M}$ slices we see that most of the turbulent material inside the dense shell is subsonic at all times.  It is important to point out here that the realistic adiabatic equation of state used in these simulations is critical for the correct calculation of the sonic Mach number. In our simulations we are able to follow thermodynamics of the gas and hence the temperature and Mach number fluctuations.

In the time series of the mass-weighted density projections we ignore the ambient material by only including gas with $T>10$ K. These density projections show gas configurations similar to the density slice plots. In the first panel at $t=1.7$ kyr we can see the locations of high-density shells that have not completely merged. These correspond to the dark, higher-density curves on the top right. As the simulation progresses the high-density shells of individual bubbles merge and the gas becomes more homogeneous within the cluster shell.

In Figure \ref{fig:density_PDF_timeseries} we show the PDFs for density, temperature and Mach number for different snapshots. The narrow peaks at $\rho \sim 10^{-21} \rm \ g \ cm^{-3}$, T $ \sim 10 \ \rm K$, and the peaks between $ \mathcal{M} \sim 10^{-4} $--$10^{-2}$ correspond to the ambient material. The density PDFs (left panel) show non-lognormal behavior throughout the evolution of the bubble.  The density PDF of subsonic and supersonic turbulence has been extensively studied and, for isothermal turbulence, takes on a lognormal form \citep{Vazquez_1994,Federrath_2008,Burkhart_2009,Kainulainen13a,Burkhart_2012,Burkhart_2015a}. 
 This is primarily attributed to the application of the central limit theorem to a hierarchical (e.g. turbulent) density field generated by a multiplicative process, such as shocks.  However, for non-isothermal turbulence or for turbulence with self-gravity (e.g. adiabatic equation of state) a lognormal is no longer observed \citep{federrath12,Collins12a,Nolan15,Mocz2017,Burkhart2018}. The density PDF is significantly affected by temperature variations (left panel of Figure \ref{fig:density_PDF_timeseries}), as expected for the case of non-isothermal gas with heating/cooling \citep{1998ApJ...504..835S,2020MNRAS.tmp.1082M}.
 The fact that the density PDF is highly non-lognormal once stellar winds become important may indicate that, for second generations of stars forming near wind-blown bubbles, star formation theories that rely on the lognormal density PDF may not be applicable. 
 
 The temperature PDF (middle panel in Figure~\ref{fig:density_PDF_timeseries}) has three main features: a peak at $T = 10$ K, a peak at $T \sim 10^{4}$ K, and a peak at $T \sim 10^7$ K. The first peak corresponds to the ambient material. The middle peak corresponds to the material just interior to bubble shell. The material below this peak value, but above 10 K, corresponds to the swept up material in the shell. This shell material begins to cool for two reasons since radiative cooling ($L_{\rm cool} \propto n^2_{\rm X} \Lambda(T,Z)$) depends on the cooling function $\Lambda(T,Z)$ and the election number density of the material $n_{\rm X}$.  First, the material reaches higher densities when it is accumulated onto the outer shell. Second, the cooling function has a local maximum of efficiency for material at $T \sim 10^4$~K at solar metallicity. Because of this we see that the shell is the only material that efficiently cools during the simulation, which we discuss in more detail in Section~\ref{sec:cool}. The final feature at $T \sim 10^7$~K corresponds to the material inside the cluster bubble. Unlike the shell material this only cools slightly. This is likely because the material inside the bubble is kept at low densities and is constantly experiencing wind-wind collisions that shock heat the material to high temperatures where the cooling function, $\Lambda(T,Z)$, is very low.

The Mach number PDF (right panel in Figure~\ref{fig:density_PDF_timeseries}) shows three features. The only sonic feature $\log_{10}\mathcal{M}~>~0$ corresponds to the bubble shell. The other feature at $\log_{10}\mathcal{M}>$ -$2$ corresponds to the collective cluster wind. These two are the dominating bubble features. As the simulation evolves we see that these features do not change but only increase in magnitude due to the larger bubble and shell volumes as the simulation evolves. The third feature between $\log_{10}\mathcal{M} \sim$ -$2$ -- -$4$ corresponds to the background material. This peak evolves towards higher Mach number at later times because, although the temperature of the background medium remains constant, the material's velocity increases slightly as it becomes gravitationally attracted to the cluster bubble.

\begin{figure*}
\includegraphics[width=1.0\linewidth]{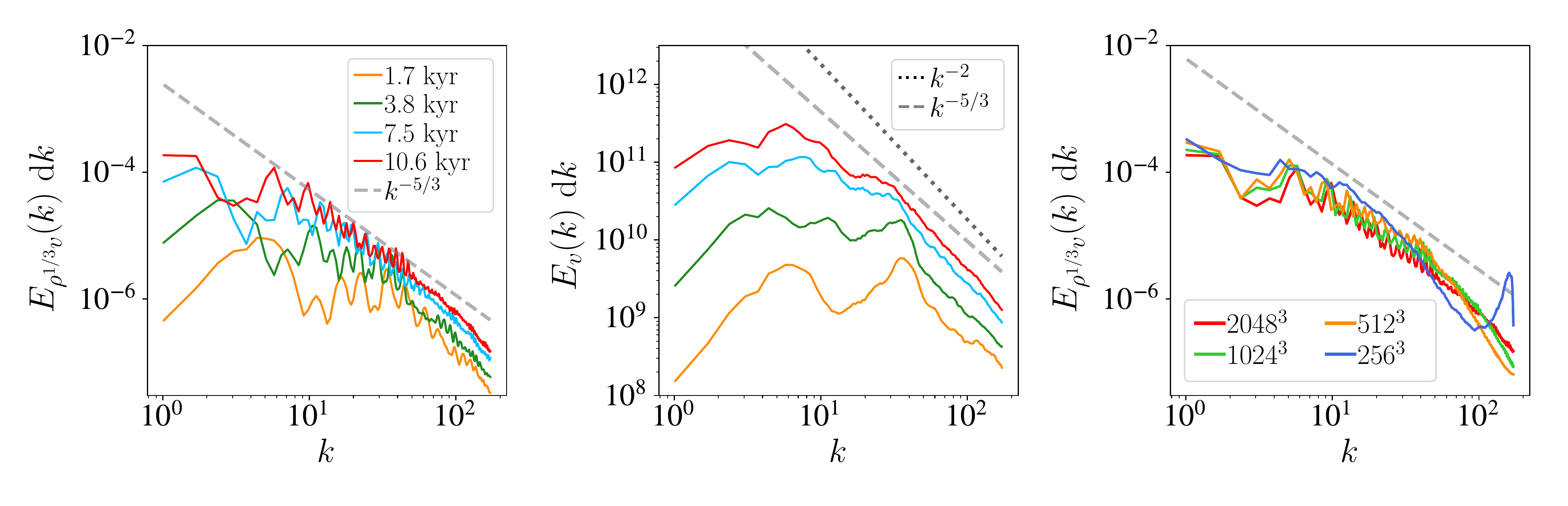}

\caption{Left: Time evolution of the density-weighted velocity power spectrum. Color corresponds to a different snapshot in the simulation. We find that it follows a Kolmogorov power spectrum (gray dashed line) Middle: Time evolution of the velocity power spectrum. The colors correspond to the same times as in the left plot. We compare this to -5/3 Kolmogorov power spectrum (gray dashed line) and -2 Burgers slope (black dotted line). Right: Resolution test for the density-weighted  velocity power spectrum. The color corresponds to the maximum levels of refinement for the AMR grid.}
\label{fig:density_weigthed_velocity}
\end{figure*}

\subsection{Turbulent Power Spectrum}\label{sec:power spectrum}
As the wind blown bubble expands into the ambient medium, turbulent density fluctuations develop in the inner bubble post-shock region. Since these fluctuations are largely subsonic (i.e. developed in the postshock gas) we may expect to find Kolmogorov-like turbulence inside the bubble.

In the limit of incompressible (sub-sonic) turbulence, density fluctuations are not relevant
and the density and kinetic energy power spectrum should evolve in a similar fashion.  Therefore for incompressible turbulence the Fourier power spectrum slope is expected to remain close to the Kolmogorov index of -5/3.
\citep{GS95,chepurnov15}.
If instead, we expected supersonic turbulence 
or a strong signature of self-gravity, the density power spectral slope would be significantly flatter than Kolmogorov \citep{Kowal07a,burkhart10,Collins12a}. 
For the velocity power spectrum, supersonic flows approach the limit of Burgers turbulence with a slope of -2. 
As for the power spectrum of density, shocks can create small-scale density enhancements (e.g.,
\cite{2005ApJ...624L..93B}; \cite{2007ApJ...666L..69K}, which in turn induce more power on small
scales and significantly flatten the spectral slope as compared to incompressible turbulence. \citet{Kritsuk07a}  proposed to use the density-weighted velocity power spectrum,  $u\equiv \rho^{1/3} v$, in order to restore the Kolmogorov scaling in the power spectra and second order structure function in compressible high Mach number hydrodynamic turbulence. 

Following \citet{Kritsuk07a}, we show the density-weighted velocity power spectrum $u\equiv \rho^{1/3} v$ (left panel)  and the velocity power spectrum (middle panel) in Figure~\ref{fig:density_weigthed_velocity}. Different time snapshots are represented with different colors, where red represents the most advance snapshot and yellow shows the earliest snapshot. The straight gray dashed line shows the -5/3 Kolmogorov prediction and the black dotted line shows the predicted slope for Burgers' turbulence.  

The different colored lines show how the power spectrum evolves in time in our simulations. At early times (i.e., yellow line), the wind-blown bubble is expanding from the smallest scales to larger scales and the velocity power spectrum is not a well defined power law. However, as time increases a power law like feature forms with a slope consistent with a value between -2 and -5/3. In particular, for the most advanced time snapshot, we find that  the density-weighted velocity power spectrum (left panel) follows a Kolmogorov scaling  and the velocity power spectrum (middle panel) in our simulations obeys a Kolmogorov/Burgers scaling after 10 kyrs.  Our results agree with previous studies of turbulence driven by stellar winds, such as \citet{Offner2015}, who find that wind-blown bubbles affect the velocity power spectrum. 

We also perform a convergence study to determine how the power spectrum depends on the AMR grid resolution in the right panel of Figure \ref{fig:density_weigthed_velocity}.
The effective grid resolution including AMR is 2048$^3$ and the grid resolution for the power spectrum calculation is $512^3$.  From the right panel of Figure \ref{fig:density_weigthed_velocity}, the spectral slope in the range of $k\sim10$--$20$ seems to be converged for resolution greater than $512^3$. 
\citet{Kritsuk07a} suggested that the power spectral scaling  based on their time-averaged statistics from a $1024^3$ driven turbulence 
simulation may begin to  correspond to R$_e \rightarrow \infty$.  Similarly, we find our simulations of wind driven turbulence also begin to converge at around this resolution.

\subsection{Shell Properties and Cooling Time}\label{sec:shell properties}

\label{sec:cool}

\begin{figure}
\includegraphics[width=1.0\linewidth]{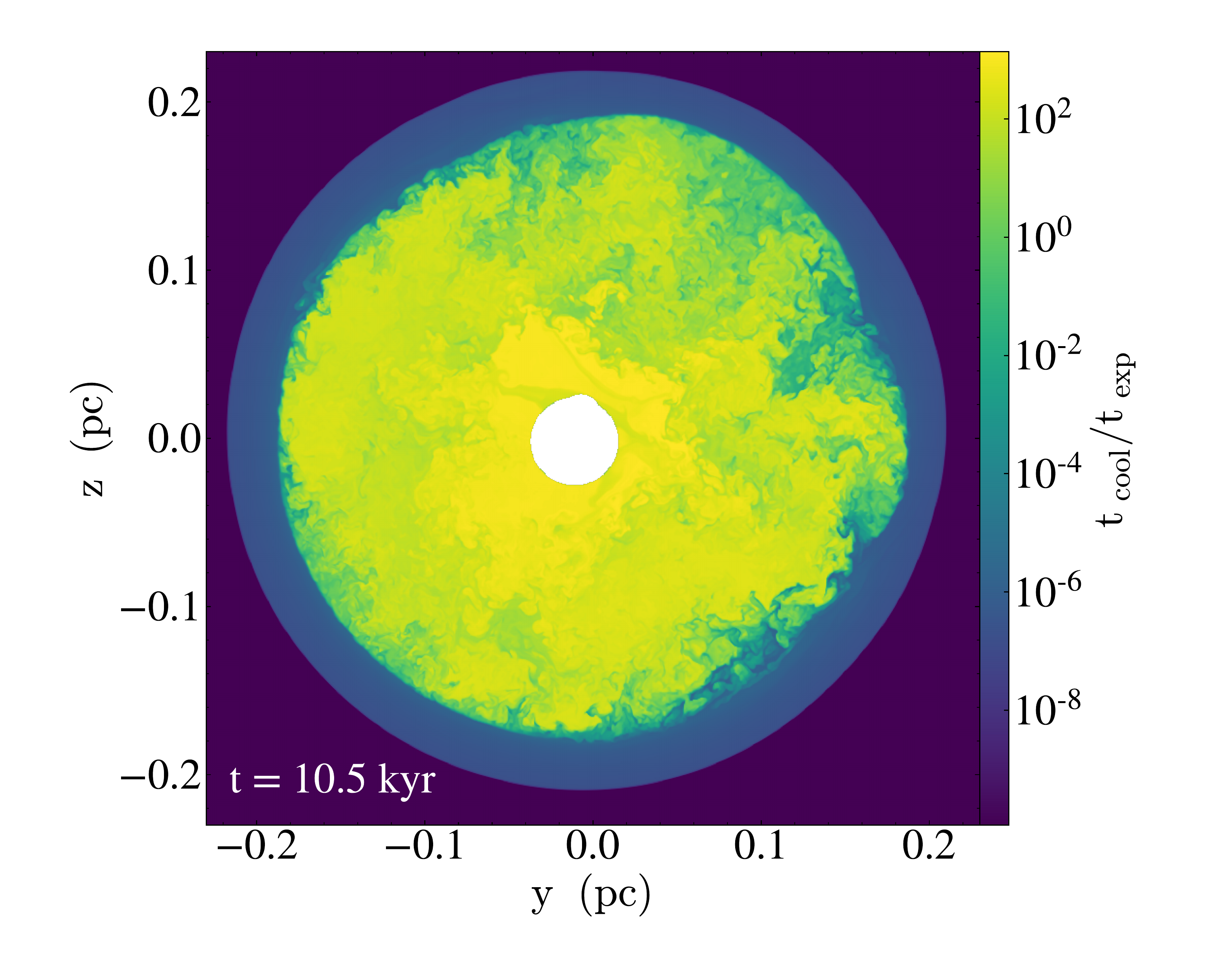}
\includegraphics[width=1.0\linewidth]{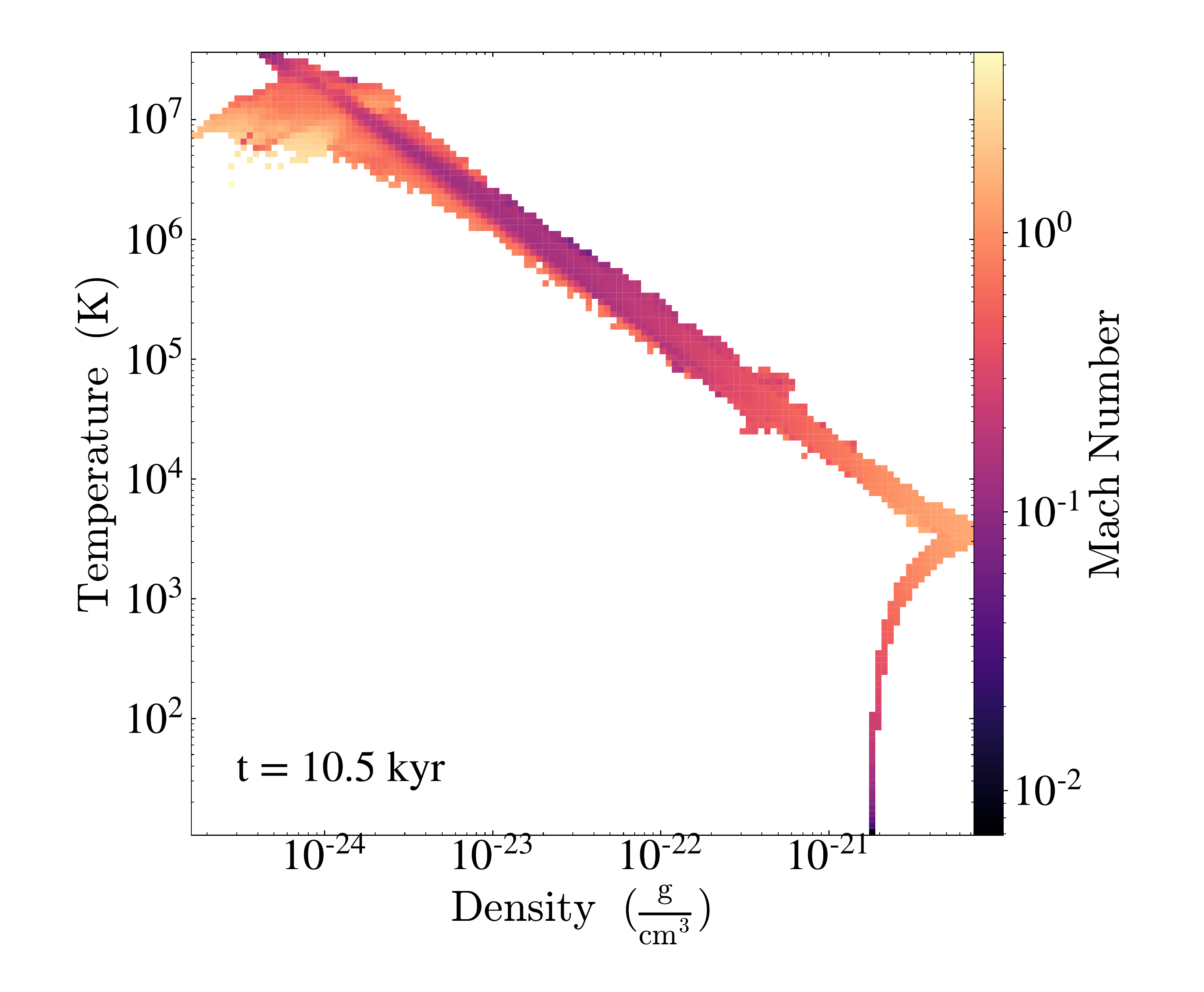}
 \centering 
\caption{Top panel: Ratio of the cooling time for the bubble material $t_{\rm cool}$  to the characteristic bubble expansion time $t_{\rm exp}$ at $t=10.5$ kyr. For these calculations we remove the cells that are dominated by numerical effects. Bottom panel: Mach number weighted by cell-mass within the cluster wind as a function of temperature and density at $t=10.5$ kyr. We show only gas with  $T > 10$ K so that we do not include contributions from the background ISM.  }
\label{fig:phaseplot}
\end{figure}

Our simulations show that after $\sim$~5~kyr the combined wind of the stars leaves the cluster itself, forming a collective cluster shell.
We calculate the shell radius as the density-weighted
average distance from the origin following equation~23 in \cite{Rosen_2017}. We obtain $R_{\rm shell} = 0.2 \ \rm pc$ at $t=10.5 \ \rm kyr$ corresponding to an approximate expansion rate $v_{\rm exp}=R_{\rm shell}/t$ of the shell to be 18.7~$\rm km \ s^{-1}$.  For comparison, we also calculate a density-weighted average velocity expansion of 9.05 $\rm km \ s^{-1}$, which is a factor of $\sim 2$ lower than the value derived previously. Typically, observations infer the bubble age from the dynamical timescale, $t_{\rm dyn} = R_{\rm shell}/v_{\rm exp}$ where both $R_{\rm shell}$ and $v_{\rm exp}$ are observationally inferred values of the shell radius and shell expansion rate, respectively \citep[e.g.,][]{2015ApJS..219...20L}. Our measured values stated above for the shell velocity imply that inferring the bubble age from these quantities will underestimate the dynamical age of the bubble since the shell velocity is decreasing in time.

In the top panel of Figure \ref{fig:phaseplot} we compare the time for the bubble material to cool, $t_{\rm cool}$, to the characteristic bubble expansion time $t_{\rm exp}=R_{\rm shell}/{v_{\rm exp}}$ using $v_{\rm exp} = 9.05 \ \rm km \ s^{-1}$ from above. In each cell we calculate the cooling time $t_{\rm cool}$ given by,
\begin{equation}
    t_{\rm cool} = \frac{E_{\rm thermal}}{{L}_{\rm cool}} = \frac{(3/2) 1.9 n_{\rm X} kT}{ 0.9 n_{\rm X}^2 \Lambda(T,Z)}.
\end{equation}
Here $E_{\rm thermal}$ is the thermal energy of the gas parcel, ${L}_{\rm cool}$ is the energy loss rate via cooling, and $\Lambda(T,Z)$ is the cooling function taken from \cite{2007ApJS..168..213G}. Here we have used $n_{\rm i} = 0.9 n_{\rm X}$ for a fully ionized plasma of solar composition, where $n_{\rm X}$ and $n_{\rm i}$ are the electron and ion number density respectively.  To calculate $ n_{\rm X} $ in each cell we use $\rho =  1.9   \mu m_{\rm p} n_{\rm X}$, where $\mu = 0.62$ assumes all helium is ionized at solar metallicity.

In the top panel of Figure \ref{fig:phaseplot} we show $t_{\rm cool} / t_{\rm exp}$ using the characteristic bubble expansion time $t_{\rm exp}~\sim~21$~ kyr in all cells. We find a cell-mass weighted average of $t_{\rm cool}~\sim~340$~kyr and a cell-mass weighted average of $t_{\rm cool} / t_{\rm exp} \sim 16$ for the hot stellar winds inside the bubble shell. The majority of the gas within the shell, especially at the center where the density is lowest, has $t_{\rm cool} \gg t_{\rm exp}$. At larger radii closer to the shell boundary we find $t_{\rm cool} \gtrsim t_{\rm exp}$. This is likely due to the turbulent mixing between the low-density, high-temperature shock-heated gas and high-density, low-temperature shell. Since $t_{\rm cool} \gg t_{\rm exp}$ the energy loss via cooling is negligible within the bubble and instead the thermalized wind energy is transferred via adiabatic expansion, thereby driving the dense bubble shell. We note that we find that the regions around the most massive stars at the center of the simulation are dominated by numerical effects and we have removed these from the calculations in Figure \ref{fig:phaseplot} as indicated by the white region. Since the majority of the gas has $t_{\rm cool} \gg t_{\rm exp}$ this region does not affect our results.

The bottom panel of Figure \ref{fig:phaseplot} shows the Mach number as a function of temperature and density at $t=10.5$ kyr within the cluster wind. We ignore the background medium by only including gas with $T > 10$ K. The low-temperature ($T \lesssim  4 \times 10^{3}$ K) and high-density material in Figure~\ref{fig:phaseplot} corresponds to the outer radius material of the bubble shell. Material at $T \sim 10^{4}$ K and mach number $\mathcal{M} > 1$ correspond to the bulk shell material. Moving upwards to higher temperature, $T = 10^4$--$10^7$~K and lower densities, $\rho = 10^{-24}$ -- $10^{-22}$~$\rm g \ cm^{-3}$, corresponds to the collective cluster winds. Most of this material is sub-sonic except at the highest temperature and lowest density.

To study the stellar wind energy of the cluster we calculate the fraction of the total energy injected by the stellar winds that go into the following: thermal energy of the hot gas interior to the shell $E_{\rm thermal} = (3/2) N k T $, the kinetic energy of the shell $E_{\rm shell} = (1/2) M_{\rm shell} v_{\rm exp}^2$, and turbulent energy of the hot gas $E_{\rm rms} = (1/2) M_{\rm B} v_{\rm rms}^2$. Throughout the simulation the energy injected by stellar winds, $L_{\rm w} = (1/2) \dot{M}_{\rm w} v_{\rm w}^2$, is $L_{\rm w} = 1.36 \times 10^{34} \ \rm erg \ s^{-1}$. At $t=3.75$ kyr the fraction of the total energy in thermal energy is $f_{\rm thermal} \approx 0.63$ and the fraction in kinetic energy of the shell is $f_{\rm shell} \approx 0.17$. We find that the fraction in turbulent energy is comparable to $f_{\rm shell}$. At $t=10.5$ kyr the results are similar except the thermal energy is slightly lower with $f_{\rm thermal} \approx 0.58$. The missing energy is likely associated with mixing of hot and cool gas and a small fraction of energy lost by cooling.


\section{Discussion and conclusions}\label{sec:discussion}

We use hydrodynamic simulations to investigate the impact of stellar winds on their environment in the first few thousand years of the wind expansion. In particular, we perform simulations that employ a realistic adiabatic equation of state in order to properly follow the adiabatic expansion of the thermalized shocked wind material that is produced by winds colliding with the winds of nearby stars and the ISM. We also include radiative cooling and self-gravity. We sample a range of mass-loss rates corresponding to intermediate and high-mass stars for a dense mass-segregated star cluster following a stochastically sampled Kroupa IMF. With these simulations, we study how stellar wind-wind collisions can drive turbulence within the expanding wind-blown bubble. We show that the turbulence driven by stellar winds is primarily subsonic likely because the shock heated material cools via adiabatic expansion and has temperatures of $\sim10^6-10^7$ K and velocities of $10^2-10^3 \; \rm{km \; s^{-1}}$.
This material continues to cool via adiabatic expansion and expands at high velocities until it interacts with the surrounding bubble shell. We find that the cluster wind material exhibits a velocity power spectrum scaling between Kolmogorov and Burgers turbulence. The power spectral scaling we observe is similar to previous numerical studies of wind driven turbulence \citep{Offner2018}. 

This is in contrast to molecular gas on larger scales in GMCs (i.e. 1-10 pc) that is predominately isothermal and cold (T $\sim$ 10 K), and has a much higher sonic Mach number.  We conclude that stellar feedback can drive small-scale turbulence from colliding stellar winds in the immediate vicinity of high-mass stars and may be able to offset dissipation of turbulent energy cascading down from larger scales.

Our study is in agreement with previous investigations that find stellar winds can drive turbulence out to about a 1 pc scale (e.g., the low $k$ bump in the power spectrum in Figure \ref{fig:density_weigthed_velocity}) but are likely not to be the dominate driver on larger scales in GMCs \citep{2006ApJ...653..416C,2010ApJ...709...27W,Offner2015,Offner2018}.  For sub-parsec length-scales in the vicinity of intermediate mass star clusters and, for the timescales we consider ($\approx 10$ kyr), stellar winds may be the dominate source of \textit{local} turbulent motions.  This is because turbulence from a larger scale cascade will be damped to velocities below 1  $\rm km \ s^{-1}$  \citep{1981MNRAS.194..809L,2015ApJ...805..118B,2018ApJ...864..116Q} while the expansion speed of our bubble is in excess of 9 $\rm km \ s^{-1}$ after 10 kyr and because timescales we consider are too short for the stars to go supernova.  The winds deposit energy and momentum into the shell but the shell wake exhibits significant non-local perturbations caused by instabilities that develop within the collective cluster wind and mixing between the cool shell gas and hot gas within the bubble. This contributes to an evolving turbulent cascade, as is evident from the velocity power spectrum.  

Our results imply that wind driven turbulence around intermediate and high-mass stars may trigger subsequent generations of ongoing star formation in the few parsec vicinity of the cluster that impinge on the shell. This second generation of stars might form in a very different manner than the first generation that is produced by cold collapsing gas from the natal molecular environment.  The density distribution is largely a power-law formed during the initial phases of star formation \citep{Kainulainen13a,Burkhart2018}. In this work, we find that the density distribution within the bubble is not lognormal due to the non-isothermal nature of the gas and a powerlaw is likely not present due to the fact that self-gravity is not important in the expanding region of the bubble, at least for the short simulation time presented here.  The bubble gas is too hot to collapse and form stars in the vicinity of the cluster and the next generation of stars likely would form further away from the cluster, triggered by the compression of the shell if the shell is able to sweep up enough gas and cool efficiently \citep{10.1093/mnras/stu1571}.

Our results should also provide an understanding of the types of environments likely to be relevant for assessing supernova  feedback in star clusters. For instance, the  dynamics of supernova in turbulent medium \citep{Martizzi2015,KimOstriker2015,Zhang2019} can be significantly different than in a homogeneous medium \citep{1977McKee}. What is more, clustering of supernova might occur before the star cluster dissolves, which  could potentially enhanced supernova feedback \citep[e.g.,][]{Kim2017,Fielding2018,Gentry2019,Karpov2020}.

Finally, our results have interesting implications for studies of clustered astrospheres in the ISM. In particular, wind blown bubbles from clusters of intermediate mass stars may alter the dynamics of cosmic ray propagation and diffusion as compared to stars in isolation. Galactic cosmic rays passing through large cavities will have their spectra efficiently cooled and thus bubbles can give rise to small-scale anisotropies in the direction to the observer \citep{Scherer_2015}. 

Future numerical studies of star cluster winds should consider including magnetic fields, in addition to the adiabatic equation of state, in order to quantitatively connect to cosmic ray observables.

Our main results from this study are as follows:
\begin{enumerate}
    \item Wind-wind collisions from the winds of intermediate and high-mass stars  drives primarily subsonic turbulence.  The turbulence exhibits a power spectrum between Kolmogorov and Burgers turbulence developed within the bubble on the bubble expansion time.
    \item An adiabatic equation of state, heating, and radiative cooling are all important effects to include when treating the physics of wind blown bubbles. An adiabatic equation of state enables a more realistic treatment of the kinetic energy injection from the fast stellar winds and the adiabatic expansion of the bubble.
    \item Dense and low-temperature shells can be a potential site for star formation if the mass accumulation is an on-going process. 
    \item  We find that the majority of the injected wind energy is in the form of thermal energy of the hot, low-density gas within the bubble and that the shell and turbulent kinentic energy within the bubble are similar throughout the simulation. As the simulation evolves, the thermal energy within the bubble decreases slightly due to adiabatic expansion, radiative cooling, and mixing of cold and hot material near the bubble shell.
\end{enumerate}


\acknowledgments
M.G-G. is grateful for the support from the Ford Foundation Predoctoral Fellowship.
 B.B. is grateful for discussions with the SMAUG collaboration.
 B.B. gratefully acknowledges generous support from the Simons Foundation Center for Computational Astrophysics (CCA) and the Harvard-Smithsonian Center for Astrophysics (CfA) Institute for Theory and Computation (ITC) postdoctoral fellowship.  A.L.R. acknowledges support from NASA through Einstein Postdoctoral Fellowship grant number PF7-180166 awarded by the \textit{Chandra} X-ray Center, which is operated by the Smithsonian Astrophysical Observatory for NASA under contract NAS8-03060. J.P.N. acknowledges support from the Institute for Theory and Computation (ITC) postdoctoral fellowship and the National Science Foundation Astronomy and Astrophysics postdoctoral fellowship award number 1402480. E.R.-R. thanks  the Heising-Simons Foundation and the Danish National Research Foundation (DNRF132) for support. A.L.R. would also like to thank her ``coworker,"  Nova Rosen, for ``insightful conversations" and unwavering support at home while this paper was being written during the Coronavirus pandemic of 2020. Her contributions were not sufficient to warrant co-authorship due to excessive napping.\footnote{Because she is a cat.}

\bibliography{main}
\bibliographystyle{aasjournal}


\end{document}